\documentclass[aps,prl,showpacs,twocolumn,amsmath,amssymb]{revtex4-1}

\usepackage{graphicx}% Include figure files
\usepackage{bm}% bold math
\usepackage{color}
\usepackage[left]{lineno}

\newcommand{\cev}[1]{\reflectbox{\ensuremath{\vec{\reflectbox{\ensuremath{#1}}}}}}

\begin{document}

%\title{Direct photoassociation of pendular butterfly Rydberg molecules}
\title{Observation of pendular butterfly Rydberg molecules}
\author{Thomas Niederpr\"um$^1$}
\author{Oliver Thomas$^{1,3}$}
\author{Tanita Eichert$^1$}
\author{Carsten Lippe$^1$}
\author{Jes\'us P\'erez-R\'ios$^2$}
\author{Chris H. Greene$^2$}
\author{Herwig Ott$^1$}
\email{ott@physik.uni-kl.de}
\affiliation{$^1$Research Center OPTIMAS, Technische Universit\"at Kaiserslautern, 67663 Kaiserslautern, Germany}
\affiliation{$^2$Department of Physics and Astronomy, Purdue University, West Lafayette, Indiana, 47907, USA}
\affiliation{$^3$Graduate School Materials Science in Mainz, Staudinger Weg 9, 55128 Mainz, Germany}

%\date{\today}

\maketitle
%introduction

{\bf Obtaining full control over the internal and external quantum states of molecules is the central goal of ultracold chemistry \cite{Carr2009} and allows for the study of coherent molecular dynamics, collisions \cite{Quemener-2012,JPR-2014,Kendrick-2015,Hauser2015} and tests of fundamental laws of physics \cite{Baron2014}. When the molecules additionally have a permanent electric dipole moment, the study of dipolar quantum gases and spin-systems with long-range interactions \cite{Baranov2012,Kadau-2016} as well as applications in quantum information processing \cite{DeMille2002} are possible. Rydberg molecules constitute a class of exotic molecules, which are bound by the interaction between the Rydberg electron and the ground state atom. They exhibit extreme bond lengths of hundreds of Bohr radii \cite{Greene2000, Bendkowsky2009} and giant permanent dipole moments in the kilo-Debye range \cite{Booth2015}. A special type with exceptional properties are the so-called butterfly molecules \cite{Hamilton2002}, whose electron density resembles the shape of a butterfly.
Here, we report on the photoassociation of butterfly Rydberg molecules and their orientation in a weak electric field.
Starting from a Bose-Einstein condensate of rubidium atoms, we fully control the external degrees of freedom and spectroscopically resolve the rotational structure and the emerging pendular states in an external electric field.
This not only allows us to extract the bond length, the dipole moment and the angular momentum of the molecule but also to deterministically create molecules with a tunable bond length and orientation. We anticipate direct applications in quantum chemistry, many-body quantum physics and quantum information processing \cite{DeMille2002,Rabl-2006, Wei2011}. }\\

\begin{figure}
\begin{center}
\includegraphics[width=0.8\columnwidth]{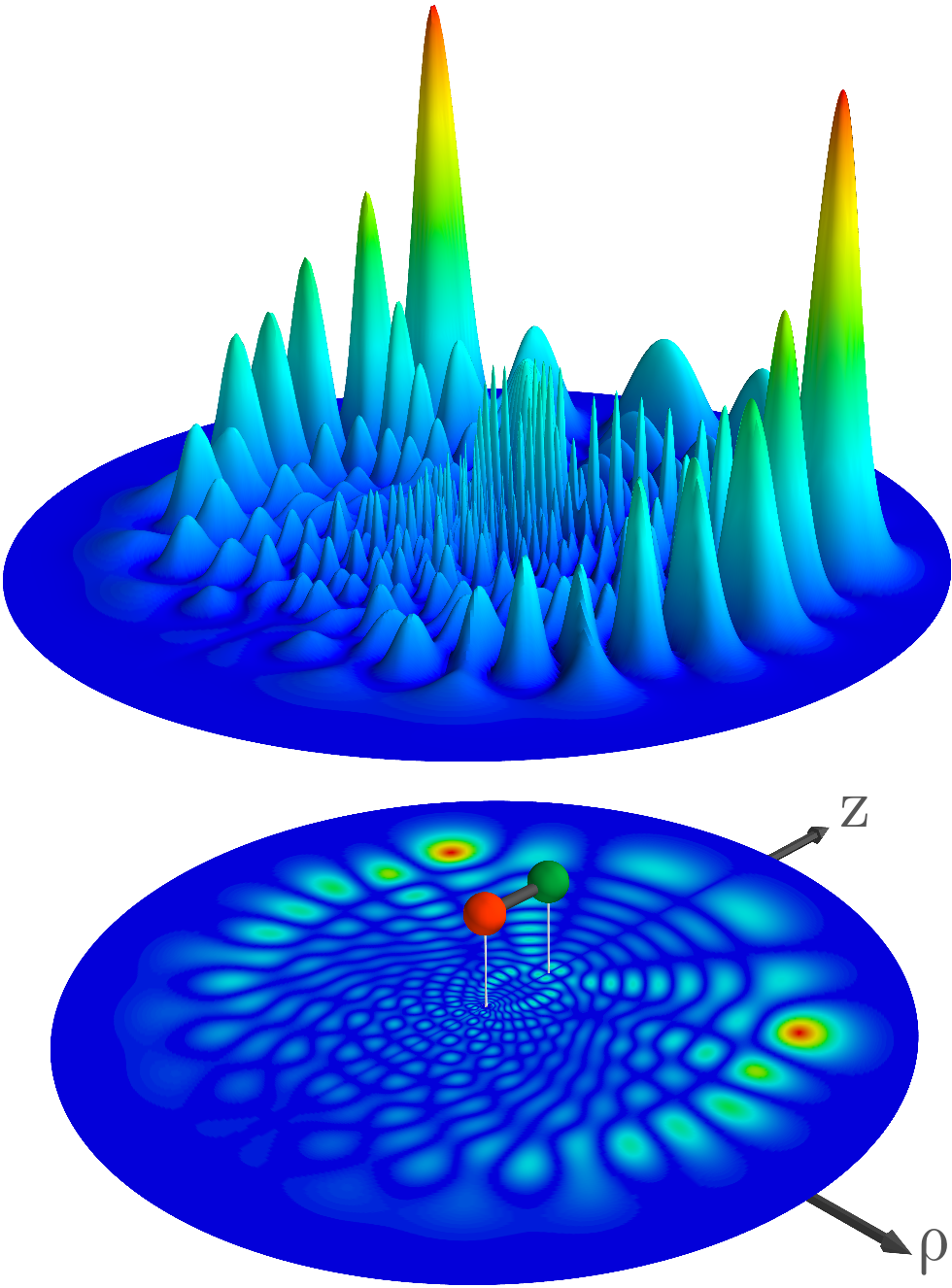}
\end{center}
\caption{Radial electron density $\rho |\Psi(z,\rho)|^2$ for a butterfly molecule on the red side of the $25P$ state of rubidium: surface plot (upper graph) and 2D projection (lower graph). A sketch of the molecule above the projection plane shows, where the $Rb^+$ ion (red) and the ground state perturber (green) are located. The bond length is $245\,\mathrm{a_0}$.} 
\label{fig:butterflyStates}
\end{figure}

\begin{figure*}[t]
\begin{center}
\includegraphics[width=\textwidth]{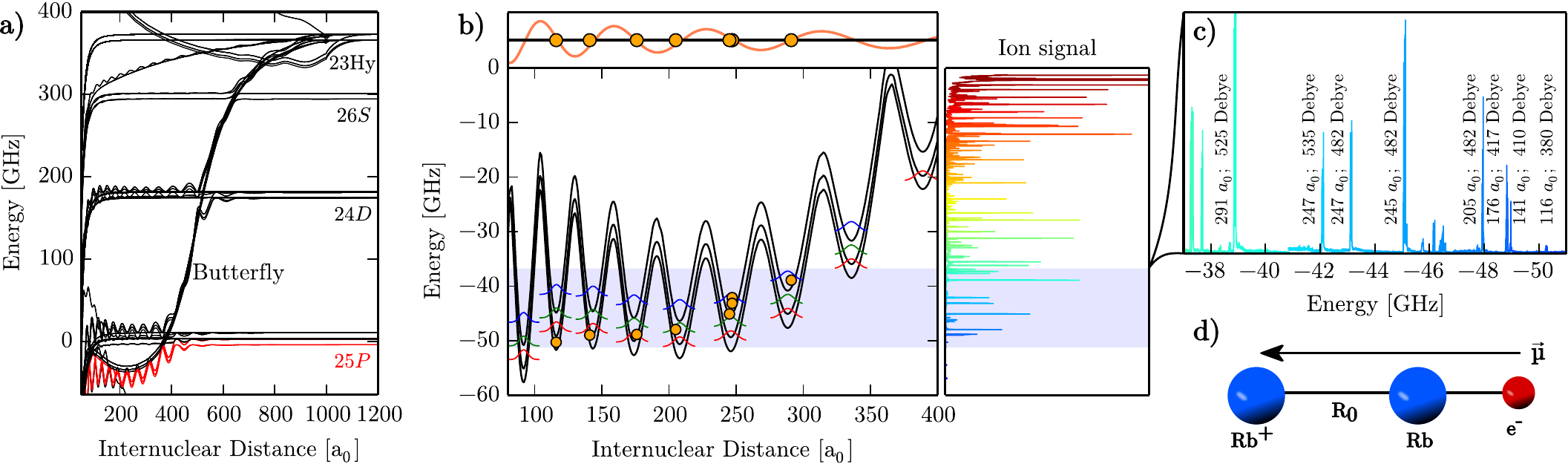}
\end{center}
\caption{(a) Adiabatic potential energy curves (PEC) for the Rydberg-ground state interaction in the vicinity of the $25P$ state. The butterfly state detaches from the n=23 hydrogenic manifold and crosses the lower lying $26S$, $24D$ and $25P$ state. A zoomed-in inner region of the lowest butterfly potentials (red lines) is shown in (b). Three different PEC (black lines) emerge, corresponding to different projections of the total angular momentum on the internuclear axis. The oscillatory shape provides a set of harmonic potential wells at different bond lengths. The lowest bound states for each PEC are sketched in red, green and blue. The orange points represent the experimentally obtained values for the binding energy and the bond length of the eight molecular states studied in detail (see text). The extension of the points denotes the error. The top panel shows how the measured bond lengths (orange points) coincide with the nodes of the $25P_{1/2}$, $m_j=1/2$ radial wavefunction (light red). The right panel shows the experimental spectrum. The color code is a guide to the eye. The zoom on the lowest energy peaks of the experimental spectrum (c) shows the determined bond length and the dipole moment of the respective peak. (d) To scale sketch of the molecule constituents averaged z-positions for a butterfly state with a bond length of $116\,\mathrm{a_0}$. The average position of the electron is located beyond the perturber atom.} 
\label{fig:Spectrum}
\end{figure*}

%Introduction to Rydberg molecules and theoretical foundation
The discovery of Rydberg molecules \cite{Bendkowsky2009} has revealed a new chemical binding mechanism. It is based on the elastic $s$- and $p$-wave scattering between the electron in a Rydberg state and a ground state perturber atom, which is located inside the electron wave function. The interaction is described by a Fermi-type contact interaction with an energy dependent scattering length \cite{Fermi,Omont,Greene2000,Khuskivadze-2002}. The bond length of these molecules is connected to the extension of the electron wave function and can reach hundreds of nanometers. In the extreme case, the interaction with the ground state atom can substantially displace the electron from the ionic core, giving rise to permanent electric dipole moments up to one kilo-Debye \cite{Booth2015}. This holds even in the case of homonuclear molecules due to the vanishing exchange energy between the constituting atoms. The extreme properties of Rydberg molecules make them versatile objects for the study of low-energy electron-atom scattering, precision calculation of potential energy curves \cite{Schlag-2016}, molecular dipole-dipole interaction, Rydberg blockade and anti-blockade effects in molecular samples as well as complex molecular dynamics.

Butterfly Rydberg molecules are dominated by the $p$-wave scattering between the Rydberg electron and the ground state atom, which maximizes the gradient of the electron wave function at the position of the ground state atom (see Methods). As a consequence, all angular momentum states of the Rydberg electron are mixed and the electron density acquires a characteristic density distribution which resembles a butterfly (Fig.\,1). The calculation of the corresponding potential energy curves (PEC) in Born-Oppenheimer approximation requires the diagonalization of the full Hamiltonian, which 
in our approach includes the fine structure in the Rydberg atom, the hyperfine interaction in the perturber atom and all $s$- and $p$-wave scattering processes (see Methods). The PECs in the vicinity of the $25P$ state are shown in  Fig.~\ref{fig:Spectrum}a. The butterfly states detach from the $n=23$ hydrogenic manifold and cross all lower lying states up to the $25P$ state. The red PECs are relevant for this work and the bound states in the strongly oscillating part of the PEC are the butterfly molecules under investigation. The electron wave function of these PECs possess a finite admixture of the $25P$-state, which we use to excite the butterfly molecules with a single photon transition.

%Experiment
The experiment was performed in a Bose-Einstein condensate (BEC) of $^{87}\mathrm{Rb}$ with a peak particle density of $4 \times 10^{14}\,\mathrm{cm^{-3}}$, an atom number of $2\times10^5$ and a temperature of $100\,\mathrm{nK}$.
All spin states of the $5S_{1/2}$, F=1 ground state were populated.
The photoassociation of the butterfly molecules was achieved using a single photon transition to the energetic vicinity of the adiabatic free two-particle state $25P_{1/2} \otimes 5S_{1/2}, {F=1}$, which serves as reference point for our spectroscopy throughout this work.
Since the created butterfly molecules can decay into ions by photoionization and associative ionization \cite{Niederpruem2015}, the produced ions are used as a probe for the creation of the molecule.

The experimental sequence consists of $500\,\mathrm{ms}$ continuous excitation and ion detection.
By detuning the excitation laser up to $-60\,\mathrm{GHz}$ in steps of $2\,\mathrm{MHz}$ we obtain the spectrum shown to the right of Fig.~\ref{fig:Spectrum}b.
%Close to the $25P$ states we observe ultralong-range Rydberg molecules \cite{Bendkowsky2009}, that are bound by hundreds of MHz.
As the laser detuning is gradually increased, we start to probe the bound states in the butterfly potential. We observe a plenitude of molecular states up to an energy of $-50\,\mathrm{GHz}$. This is in accordance with the calculated PEC. Moreover, the density of molecular states drops at detunings below $-40\,\mathrm{GHz}$, which marks the transition to a spectral region, where only the lowest bound states in each potential well are populated. The spectroscopic results directly prove the existence of butterfly molecules.

The following focuses on the ground state molecules in each potential well and demonstrates, how butterfly molecules with selected bond length and high degree of orientation can be created.

\begin{figure*}[t]
\begin{center}
\includegraphics[width=\textwidth]{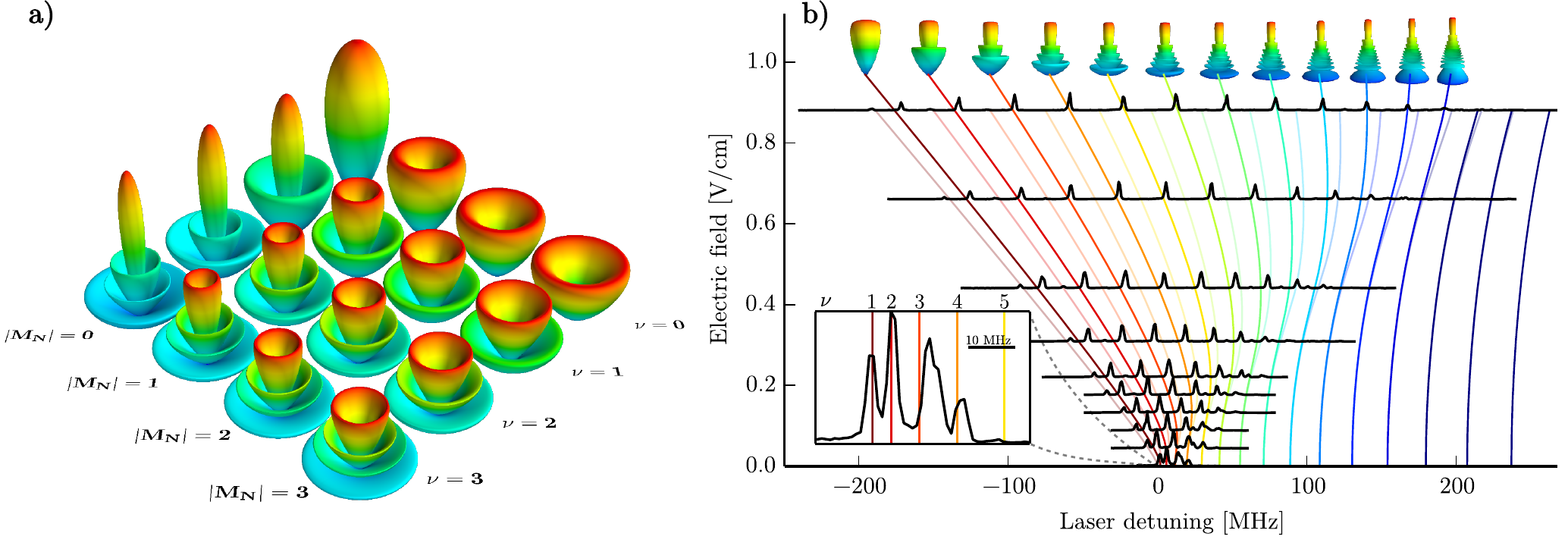}
\end{center}
\caption{(a) Plot of the orbitals $|\Phi(\theta, \phi)|$ for pendular states $|\nu,M_N\rangle$ of a dipolar molecule with $\mu= 482\,\mathrm{Debye}$ in an electric field of $1\,\mathrm{V/cm}$. (b) Spectroscopy of the butterfly state at $-47.9\,\mathrm{GHz}$ in different  electric fields. Each spectrum is normalized and shifted on the y-axis according to the applied electric field. The colored lines show the eigenenergies of a rigid rotor with a permanent dipole moment in an electric field (Eq.\,1) for $|M_{N}| = 0$ (weak lines) and $|M_{N}| = 1$ (strong lines). The dipole moment and the bond length are adapted to fit the experimental data. In this case, we find $R_{0} = 205\,\mathrm{a_0}$ and $\mu = 482\,\mathrm{Debye}$. The inset shows a zoom of the low-field ($\approx 30\,\mathrm{mV/cm}$) measurement together with the calculated position of the first five pendular states with $|M_N=1|$. We attribute the appearance of higher $\nu$ states to the mixing of the $N$ states in the residual field (Supplementary Material).} 
\label{fig:EfieldSpectrum}
\end{figure*}

Due to the symmetry breaking caused by the localized valence electron of the perturber, butterfly molecules, along with the related class of so-called trilobite molecules are the only known homonuclear molecules with a permanent electric dipole moment \cite{Li2011,Greene2000,Booth2015,Khuskivadze-2002}.
The interaction of the dipole moment $\vec{\mu}$ with an external electric field $\vec{F}$ significantly changes the rotational structure of the molecule, since the eigenstates (energy $E$) need to fulfill

\begin{align}
\left( B_e \hat{N}^2 - \vec{\mu}\cdot\vec{F}-E \right) \left| \Phi \right\rangle = 0.
\end{align}
\noindent
For vanishing electric field, the molecule behaves as  a rigid rotor of length $R_{0}$ and reduced mass $m_r=m_{Rb}/2$ using the rotational quantum number $N$, its projection on the molecular axis $M_{N}$ and the rotational constant $B_e = \hbar^2/2I$, where $I=m_{r} R_{0}^2$ represents the moment of inertia of the rigid rotor.
When, on the other hand, the interaction with the electric field dominates the rotational constant ($\omega = |\vec{\mu}\cdot\vec{F}|/B_e \gg 1$), the molecule enters so-called pendular states \cite{Rost1992, Block1992}, which are characterized by the quantum number $\nu$ and differ in the degree of orientation of the molecule with respect to the field axis (Fig.~\ref{fig:EfieldSpectrum}a).
In this regime only the angular momentum projection $M_{N}$ remains a good quantum number, while the rotational quantum numbers $N$ are strongly mixed.

In Fig.~\ref{fig:EfieldSpectrum}b, we show the electric field-dependent spectra of the butterfly state at $-47.9\,\mathrm{GHz}$ along with the fitted theory of Eq.\,1. Due to the huge dipole moments of the butterfly molecules, fields of $1\,\mathrm{V/cm}$ are sufficient to put the system deep in the pendular regime. Varying the electric field, we can directly observe the crossover from rotational states to pendular states. At fields of $4\,\mathrm{V/cm}$, the degree of orientation $\omega$ exceeds previous record values \cite{Block1992} by a factor of two. Since we photoassociate the butterfly molecules in a Bose-Einstein condensate, which has no angular momentum in the center of mass motion, only states with $|M_{N}|=0$ and $|M_{N}|=1$ can be observed (Supplementary Material). This leads to a very clean spectrum of pendular states, which, in contrast to previous studies \cite{Bendkowsky2009,Li2011,Booth2015,Bellos-2013,Greene-2006}, allows for the precise determination of the dipole moment $\mu$ and the bond length $R_{0}$.

Applying this method to eight of the lowest spectroscopic lines in the butterfly spectrum, we can extract the respective dipole moments and bond lengths (Fig.~\ref{fig:Spectrum}c).
With this additional information we can improve our interpretation of the butterfly spectrum and compare it to theory.
This is demonstrated in Fig.~\ref{fig:Spectrum} b), where the orange points mark the measured bond lengths and binding energies for the eight studied lines.
It is remarkable how well the measured bond lengths agree with the wells of the model potential.
In fact, the positions of the wells are mainly determined by the strong gradient at the radial nodes of the $p$-state wavefunction and thus the bond length of the butterfly molecules is an indirect measure for the position of the $25P$ radial nodes. A comparison of the measured bond lengths with the nodes of the calculated $25P_{1/2}$ wavefunction is shown in the top panel of Fig.~\ref{fig:Spectrum}b).

Butterfly molecules can also exhibit counter-intuitive properties. In a semiclassical picture of molecules with one valence electron, the average position of the electron is always located between the two nuclei. The maximum possible dipole moment is thus limited by the bond length, as realized, e.g., in an ionic bond. The strong interference of the electron caused by the scattering with the ground state atom (Fig.\,1) permits the butterfly molecules to exceed this limit and locate the electron's average position beyond the perturber (Fig.~\ref{fig:Spectrum} d).
This effect is seen experimentally for the deepest bound butterfly molecules.
The state at $-50.3\,\mathrm{GHz}$ possesses a dipole moment of $380\,\mathrm{Debye}$, which is 30\% larger than that of a single charge separated by the measured bond length of $116\,\mathrm{a_0}$. This phenomenon is in agreement with theoretical calculations of the dipole moment.

Due to the deep potential wells (Fig.~\ref{fig:Spectrum}b) inward tunneling and subsequent decay should not limit the lifetime of the studied butterfly molecules. In a separate experiment using short excitation pulses (see Methods) we were able to determine the lifetimes of the strongest observed butterfly molecules. For low densities the lifetime of the molecules is on the order of $20\,\mathrm{\mu s}$, which is compatible with the lifetime of the $25P$ Rydberg state. 

The long lifetime along with the negligible motion of the ground state atoms in the BEC enables narrow photoassociation lines and thus a clean optical spectroscopy. This allows for the deterministic creation of vibrational ground state molecules with a selected bond length and dipole moment. In a weak electric field we can furthermore restrict the rotation of the molecule and precisely set the orientation of the molecule with respect to the field axis. This high degree of control over molecules with giant dipole moments even in a sample with only one atomic species opens many perspectives for future applications. The high precision of the molecular spectroscopy can serve as a benchmark for quantum chemical calculations of molecular states. Rydberg molecules are promising candidates to implement dipolar interactions in many-body quantum systems. The comparable energy scale of the rotational structure and the dipole-dipole interaction is unique in the world of polar molecules and will complement studies using heteronuclear alkali dimers in the rovibrational ground state \cite{Baranov2012}. Starting from pre-associated weakly bound ground state molecules in a 3D optical lattice, spin physics with pendular states are feasible \cite{Baranov2012}. Dressing weakly bound molecules with a Rydberg molecule might even allow to study coherent tunneling dynamics with long-range interactions. It will be interesting in the future to extend the concepts of the Rydberg blockade and anti-blockade to interacting Rydberg molecules. Butterfly molecules might also be a good starting point to prepare heavy Rydberg systems \cite{Reinhold2005}. Ultimately pendular states of dipolar molecules could also be a building block to scalable quantum information processing \cite{DeMille2002, Wei2011}.
%and be a good starting point to study heavy Rydberg systems\cite{Reinhold2005}. 

\subsubsection*{METHODS}

{\bf Theoretical model}\\
%The interaction of a Rydberg atom with a neutral atom located inside the classical Rydberg orbit can be modeled by means of the quasi-free electron picture, developed by Fermi~\cite{Fermi} in order to explain the shifts of Rydberg atoms in a buffer gas \cite{Amaldi}.
%The original idea of Fermi was to describe the electron-neutral interaction as a delta-function potential proportional to the scattering length, the so-called Fermi pseudopotential.
%Fermi's approach turns out to be adequate for low energy electrons, {\it i.e.}, highly excited Rydberg states.
%However, in some systems the electron can be momentarily trapped due to the existences of shape resonances, as in $^{87}$Rb which is known to have  a shape resonance in the $^{3}$P$^{0}$ symmetry for $e^{-}$ -Rb(5s)  scattering at 0.02 eV.
%Our calculations utilize the Omont \cite{Omont} $p$-wave  pseudo potential to incorporate the $p$-wave shape resonance.
%The present calculation  also includes the fine structure effects of  the Rydberg atom, as well as the hyperfine structure of the perturbing atom  $^{87}$Rb 5$S_{1/2}$. 
Rich spectral features, such as average line shifts and collisional broadening, arise when a Rydberg atom is excited in a high density gas. These effects have been understood since Fermi's idea of a quasi-free electron\cite{Fermi} could successfully explain the early experiments by Amaldi and Segr\`e \cite{Amaldi}. Extending Fermi's work, Omont later generalized the zero-range $s$-wave potential to higher partial waves\cite{Omont}. The contribution of the $p$-wave becomes particularly important in alkali atoms, as those show a low energy shape resonance in the $^3P^o$ scattering channel. In order to model the high resolution spectroscopy of rovibrational diatomic Rydberg levels, we not only include the singlet electron-Rb $s$-wave and $p$-wave scattering information but also the spin-orbit splitting of the Rydberg state and the ground state $^{87}\mathrm{Rb}$(5$S_{1/2}$) hyperfine structure.​ Thus, the Hamiltonian for the Rydberg-perturber interaction in atomic units reads as \cite{Anderson2014}

 \begin{align}
\hat{H} &= \hat{H}_0 \nonumber\\
&+ 2\pi \left[a_s^S(k_R)\hat{\mathbb{I}}^S + a_s^T(k_R)\hat{\mathbb{I}}^T\right]\delta^{(3)}(\vec{r}-\vec{R}) \nonumber \\
&+ 6\pi \left[a_p^S(k_R)\hat{\mathbb{I}}^S + a_p^T(k_R)\hat{\mathbb{I}}^T\right]\delta^{(3)}(\vec{r}-\vec{R})\frac{\cev{\nabla}\cdot\vec{\nabla}}{k_R^2} \nonumber \\
&+ A\hat{\vec{S}}_2\cdot\hat{\vec{I}}_2,
\end{align}

\noindent
where the position of the Rydberg electron is denoted as $\vec{r}$, whereas the position of the perturber 
from the Rydberg core is expressed as $\vec{R}$. $\hat{H}_0$ represents the unperturbed Rydberg states involved in 
the calculations, accounting for the appropriate quantum defects \cite{Lorenzen-1983}. $a_s^S(k_R)$ and $a_s^T(k_R)$ denotes the singlet 
and triplet $s$-wave $e^{-}$ -Rb(5s) scattering lengths, and analogously $a_p^S(k_R)$ and $a_p^T(k_R)$
 represent the singlet and triplet $p$-wave $e^{-}$ -Rb(5s) scattering volume, respectively \cite{Khuskivadze-2002}. The momentum of 
 the Rydberg electron evaluated at the perturber position is $k_R=\sqrt{-1/n_{*}^2+2/R}$, where $n^{*}$ stands for the effective quantum number 
 of the level of interest. In Eq.(2), the triplet and singlet 
  components of the $e^{-}$ -Rb(5s) scattering lengths are associated with the projectors 
  $\hat{\mathbb{I}}^T = \hat{{\bm S}}_{1} \cdot \hat{ {\bm S}}_{2}+ 3/4$ and 
  $\hat{\mathbb{I}}^S=\hat{\mathbb{I}}-\hat{\mathbb{I}}^T$, respectively. The last term in Eq.(2) 
  represents the hyperfine structure of the perturbing atom $^{87}$Rb 5$S_{1/2}$ with $A$= 3.417\,GHz. The 
  present Hamiltonian is diagonalized in a truncated basis of atomic levels, along with the electron and nuclear
   spin degrees of freedom for the perturbing ground state atom, {\it i.e.}, $|n L_1 J_1 m_{J1} \rangle \otimes |m_{s2} m_{i2} \rangle$, where $L_1$, $J_1$ and $m_{J1}$ are associated with the Rydberg electron as in Refs. \cite{Anderson2014,Sadeghpour2013}, whereas $|m_{s2} m_{i2}\rangle$ stands for the hyperfine levels in the perturbing atom.\\

{\bf Experimental setup}\\
Starting from a 3D magneto-optical trap, a Bose-Einstein condensate of $^{87}$Rb is prepared by forced evaporation in a crossed YAG dipole trap. We reach a BEC of $2 \times 10^5$ rubidium-87 atoms at final trapping frequencies of $2\pi \times 150\,\mathrm{Hz}$ in the radial and $2\pi \times 80\,\mathrm{Hz}$ in the axial direction. The photoassociation of the butterfly molecules is done with a frequency doubled, continuous wave dye laser at a wavelength of $297\,\mathrm{nm}$ and a total power of $20\,\mathrm{mW}$, focused to a beam waist of $40\,\mathrm{\mu m}$. The excitation pulse has a duration of $500\,\mathrm{ms}$, during which the generated ions are detected with a discrete dynode electron multiplier. A small electric field of $30\,\mathrm{mV/cm}$ at the position of the BEC is needed to guide the ions to the detector and is thus always present, if not higher fields are applied. The total detection efficiency  for a single created ion is $44\%$. For the measurements presented here the linear polarization of the excitation light was orthogonal to the electric field axis, yielding strong transitions to the $M_N = 1$ states (see also Supplementary Material). 

The lifetime of the butterfly molecules was measured in a time of flight experiment, as described in \cite{Niederpruem2015}. After a short laser pulse of $1\,\mathrm{\mu s}$ the produced atomic and molecular ions are recorded. We extract the lifetime from the exponential decay of the molecular ion signal.\\

\bibliographystyle{abbrv}
\bibliography{butterfly_citation}

\begin{thebibliography}{10}

\bibitem{Amaldi}
E.~Amaldi and E.~Segr\`{e}.
\newblock Effect of pressure on high terms of alkaline spectra.
\newblock {\em Nature}, 133:141, 1934.

\bibitem{Anderson2014}
D.~A. Anderson, S.~A. Miller, and G.~Raithel.
\newblock Angular-momentum couplings in long-range {R}b$_2$ {R}ydberg
  molecules.
\newblock {\em Phys. Rev A}, 90:062518, Sept. 2014.

\bibitem{Baranov2012}
M.~A. Baranov, M.~Dalmonte, G.~Pupillo, and P.~Zoller.
\newblock Condensed matter theory of dipolar quantum gases.
\newblock {\em Chem. Rev.}, 112:5012--5061, 2012.

\bibitem{Baron2014}
J.~Baron, W.~C. Campbell, D.~DeMille, J.~M. Doyle, G.~Gabrielse, Y.~V.
  Gurevich, P.~W. Hess, N.~R. Hutzler, E.~Kirilov, I.~Kozyryev, B.~R. O'Leary,
  C.~D. Panda, M.~F. Parsons, E.~S. Petrik, B.~Spaun, A.~C. Vutha, and A.~D.
  West.
\newblock Order of magnitude smaller limit on the electric dipole moment of the
  electron.
\newblock {\em Science}, 343:269, 2014.

\bibitem{Bellos-2013}
M.~A. Bellos, R.~Carollo, J.~Banerjee, E.~E. Eyler, P.~L. Gould, and W.~C.
  Stwalley.
\newblock Excitation of weakly bound molecules to trilobitelike {R}ydberg
  states.
\newblock {\em Phys. Rev. Lett.}, 111:053001, 2013.

\bibitem{Bendkowsky2009}
V.~Bendkowsky, B.~Butscher, J.~Nipper, J.~P. Shaffer, R.~Low, and T.~Pfau.
\newblock Observation of ultralong-range {R}ydberg molecules.
\newblock {\em Nature}, 458(7241):1005--1008, 2009.

\bibitem{Block1992}
P.~A. Block, E.~J. Bohac, and R.~E. Miller.
\newblock Spectroscopy of pendular states: The use of molecular complexes in
  achieving orientation.
\newblock {\em Phys. Rev. Lett.}, 68:1303--1306, 1992.

\bibitem{Booth2015}
D.~Booth, S.~T. Rittenhouse, J.~Yang, H.~R. Sadeghpour, and J.~P. Shaffer.
\newblock Production of trilobite {R}ydberg molecule dimers with kilo-debye
  permanent electric dipole moments.
\newblock {\em Science}, 348(6230):99--102, 2015.

\bibitem{Carr2009}
L.~D. Carr, D.~DeMille, R.~V. Krems, and J.~Ye.
\newblock Cold and ultracold molecules: science, technology and applications.
\newblock {\em New J. Phys.}, 11:055049, 2009.

\bibitem{DeMille2002}
D.~DeMille.
\newblock Quantum computation with trapped polar molecules.
\newblock {\em Phys. Rev. Lett.}, 88:067901, 2002.

\bibitem{Fermi}
E.~Fermi.
\newblock Sopra lo spostamento per pressione delle righe elevate delle serie
  spettrali.
\newblock {\em Nuovo Cimento}, 11:157--164, 1934.

\bibitem{Greene2000}
C.~H. Greene, A.~S. Dickinson, and H.~R. Sadeghpour.
\newblock Creation of polar and nonpolar ultra-long-range {R}ydberg molecules.
\newblock {\em Phys. Rev. Lett.}, 85:2458--2461, Sep 2000.

\bibitem{Greene-2006}
C.~H. Greene, E.~L. Hamilton, H.~Crowell, C.~Vadla, and K.~Niemax.
\newblock Experimental verification of minima in excited long-range {R}ydberg
  states of {Rb}$_{2}$.
\newblock {\em Phys. Rev. Lett.}, 97:233002, 2006.

\bibitem{Hamilton2002}
E.~L. Hamilton, C.~H. Greene, and H.~R. Sadeghpour.
\newblock Shape-resonance-induced long-range molecular {R}ydberg states.
\newblock {\em J. Phys. B: At., Mol. Opt. Phys.}, 35(10):L199, 2002.

\bibitem{Hauser2015}
D.~Hauser, S.~Lee, F.~Carelli, S.~Spieler, O.~Lakhmanskaya, E.~S. Endres, S.~S.
  Kumar, F.~Gianturco, and R.~Wester.
\newblock Rotational state-changing cold collisions of hydroxyl ions with
  helium.
\newblock {\em Nat. Phys.}, 11:467, 2015.

\bibitem{Kadau-2016}
H.~Kadau, M.~Schmidt, M.~Wenzel, C.~Wink, T.~Maier, I.~Ferrier-Barbut, and
  T.~Pfau.
\newblock Observing the {R}osensweig instability of a quantum ferrofluid.
\newblock {\em Nature}, 530:194--197, 2016.

\bibitem{Kendrick-2015}
B.~K. Kendrick, J.~Hazra, and N.~Balakrishnan.
\newblock The geometric phase controls ultracold chemistry.
\newblock {\em Nature Communications}, 6:7918, 2015.

\bibitem{Khuskivadze-2002}
A.~A. Khuskivadze, M.~I. Chibisov, and I.~I. Fabrikant.
\newblock Adiabatic energy levels and electric dipole moments of {R}ydberg
  states of {Rb}$_{2}$ and {Cs}$_{2}$ dimers.
\newblock {\em Phys. Rev A}, 66:042709, 2002.

\bibitem{Li2011}
W.~Li, T.~Pohl, J.~M. Rost, S.~T. Rittenhouse, H.~R. Sadeghpour, J.~Nipper,
  B.~Butscher, J.~B. Balewski, V.~Bendkowsky, R.~L{\"o}w, and T.~Pfau.
\newblock A homonuclear molecule with a permanent electric dipole moment.
\newblock {\em Science}, 334(6059):1110--1114, 2011.

\bibitem{Lorenzen-1983}
C.~J. Lorenzen and K.~Niemax.
\newblock Quantum defects of the $n ^{P}_{1/2,3/2}$ levels in $^{39}${K} {I}
  and $^{85}${R}b {I}.
\newblock {\em Physica Scripta}, 27:300, 1983.

\bibitem{Niederpruem2015}
T.~Niederpr\"um, O.~Thomas, T.~Manthey, T.~M. Weber, and H.~Ott.
\newblock Giant cross section for molecular ion formation in ultracold
  {R}ydberg gases.
\newblock {\em Phys. Rev. Lett.}, 115:013003, 2015.

\bibitem{Omont}
A.~Omont.
\newblock On the theory of collisions of atoms in {R}ydberg states with neutral
  particles.
\newblock {\em J. Physique}, 38:1343, 1977.

\bibitem{JPR-2014}
J.~P\'{e}rez-R\'{i}os, M.~Leperes, R.~Vexiau, N.~Bouloufa-Maafa, and O.~Dulieu.
\newblock Progress toward ultracold chemistry: ultracold atomic and photonic
  collisions.
\newblock {\em J. Phys.: Conf. Ser.}, 488:012031, 2014.

\bibitem{Quemener-2012}
G.~Qu\'{e}m\'{e}ner and P.~S. Julienne.
\newblock Ultracold molecules under control!
\newblock {\em Chem. Rev.}, 112:4949--5011, 2012.

\bibitem{Rabl-2006}
P.~Rabl, D.~DeMille, J.~M. Doyle, M.~D. Lukin, R.~J. Schoelkopf, and P.~Zoller.
\newblock Hybrid quantum processors: Molecular ensembles as quantum memory for
  solid state circuits.
\newblock {\em Phys. Rev. Lett.}, 97:033003, 2006.

\bibitem{Reinhold2005}
E.~Reinhold and W.~Ubachs.
\newblock Heavy {R}ydberg states.
\newblock {\em Mol. Phys.}, 103(10):1329--1352, 2005.

\bibitem{Rost1992}
J.~M. Rost, J.~C. Griffin, B.~Friedrich, and D.~R. Herschbach.
\newblock Pendular states and spectra of oriented linear molecules.
\newblock {\em Phys. Rev. Lett.}, 68:1299--1302, Mar 1992.

\bibitem{Sadeghpour2013}
H.~R. Sadeghpour and S.~T. Rittenhouse.
\newblock How do ultralong-range homonuclear {R}ydberg molecules get their
  permanent dipole moments?
\newblock {\em Mol. Phys.}, 111:1902, 2013.

\bibitem{Schlag-2016}
M.~Schlagm\"{u}ller, T.~C. Leibisch, H.~Nguyen, G.~Lochead, F.~Engel,
  F.~B\"{o}ttcher, K.~M. Westphal, K.~S. Kleinbach, R.~L\"{o}w, S.~Hofferberth,
  T.~Pfau, J.~P\'{e}rez-R\'{i}os, and C.~H. Greene.
\newblock Probing and electronc scattering resonance using {R}ydberg molecules
  within a dense and ultracold gas.
\newblock {\em Phys. Rev. Lett.}, 116:053001, 2016.

\bibitem{Wei2011}
Q.~Wei, S.~Kais, B.~Friedrich, and D.~Herschbach.
\newblock Entanglement of polar molecules in pendular states.
\newblock {\em J. Chem. Phys.}, 134(12), 2011.

\end{thebibliography}

%\begin{thebibliography}{999}
%\bibitem{Jaklevic1964}
%Jaklevic, R. C., Lambe, J., Silver, A. H. \& Mercereau, J. E. Quantum Interference Effects in %Josephson Tunneling. {\it Phys. Rev. Lett.} {\bf 12}, 159-160 (1964).

%\end{thebibliography}

\vspace{0.5cm}

\subsubsection*{SUPPLEMENTARY MATERIAL}
\begin{figure}[t]
\begin{center}
\includegraphics[width=\columnwidth]{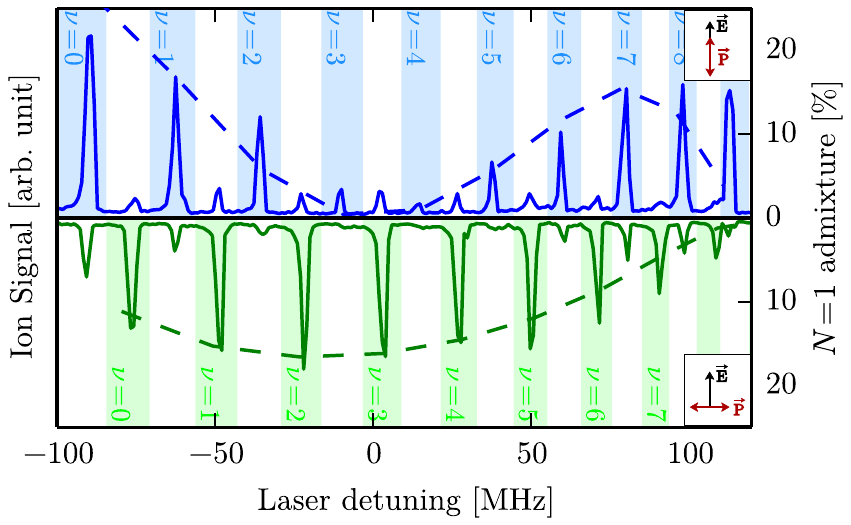}
\end{center}
\caption{(Color online). Pendular state spectrum of the butterfly state at $-47.9\,\mathrm{GHz}$ in an electric field of $0.44\,\mathrm{V/cm}$ for different angles between the linear laser polarisation and the electric field axis. While for parallel polarisation (solid blue line) mainly $|M_N|=0$ states (blue patches) are excited, we see that for orthogonal polarization (solid green line) mainly $|M_N|=1$ states (green patches) are excited. The admixture of the $N=1$ state for the individual $\nu$ is shown as dashed lines.} 
\label{fig:PolarizationDep}
\end{figure}

{\bf Polarization Dependence}\\
The simplest approximate description of the Stark effect on these Rydberg molecules that is adequate to describe the present results includes only the squared rotational angular momentum operator $\hat{N}^2$, and ignores coupling to the electronic and nuclear spin degrees of freedom. Since we photoassociate butterfly molecules from a BEC that has no angular momentum in the center of mass motion, only rotational states with $M_N=0$ and $M_N=\pm1$, corresponding to $\pi$ and $\sigma^\pm$ transitions respectively, can be excited. A comparison of the pendular state spectroscopy for different orientation of the linear laser polarization is presented in Fig.~\ref{fig:PolarizationDep}. If the laser polarization is parallel to the electric field (blue) we mainly drive a $\pi$ transition and the $|M_N| = 0$ lines dominate the spectrum. If, on the other hand, the laser polarization is orthogonal to the electric field axis (green), we drive both $\sigma^+$ and $\sigma^-$ transitions and thus mainly couple to the $|M_N| = 1$ states. 

Since the BEC provides an initial state with $N=0$, $M_N=0$ and the usual angular momentum selection rules $\Delta N = 0,\pm1$ (but $N=0 \rightarrow N'=0$ forbidden) need to be fulfilled in the photoassociation process, we can only couple to $|N=1, M_N=0,\pm1\rangle$ states. 
The coupling strength is therefore proportional to the admixture of the $N=1$ state (Fig.~\ref{fig:PolarizationDep}).
%The coupling strength therefore varies in the same way as the $N=1$ contribution to the pendular states, as demonstrated in Fig.~\ref{fig:PolarizationDep}. 

\newpage
\vspace{0.5cm}
\textbf{Acknowledgements}
\begin{acknowledgments}
We thank I. Fabrikant for providing calculated data on the non-relativistic scattering phase shifts and G. Raithel for helpful discussions. We acknowledge financial support by the DFG within the SFB/TR 49. O. T. is supported by the MAINZ graduate school. C.G. and J.P.-R. are supported in part by the U.S. National Science Foundation grant number PHY-1306905.
\end{acknowledgments}

\vspace{0.5cm}

\textbf{Author Contributions}\\
T. N., O. T., T. E. and C. L. performed the experiment.
H. O. supervised the experiment.
T. N. analyzed the data and prepared the manuscript.
T. N., O. T., J. P.-R., C. G. and H. O. developed the theoretical model.
All authors contributed to the data interpretation and manuscript preparation.

\clearpage

%##############################################################
\end{document}